\documentclass[conference]{IEEEtran}

\usepackage{algorithmic}
\usepackage{graphicx,color}
\usepackage{textcomp}
\usepackage{xcolor}
\usepackage{array}
\usepackage{multirow}
\usepackage{xurl}
\usepackage{tablefootnote}
\usepackage{amsmath,amssymb,amsfonts}
\usepackage{doi}
\usepackage{url}
\usepackage[style=ieee,doi=true]{biblatex}
\def\BibTeX{{\rm B\kern-.05em{\sc i\kern-.025em b}\kern-.08em
    T\kern-.1667em\lower.7ex\hbox{E}\kern-.125emX}}
\DeclareSourcemap{
    \maps{
        \map{ 
            \step[
                fieldsource=doi,
                match=\regexp{\{\\\_\}|\{\_\}|\\\_},
                replace=\regexp{\_}
            ]
        }
        \map{ 
            \step[
                fieldsource=url,
                match=\regexp{\{\\\_\}|\{\_\}|\\\_},
                replace=\regexp{\_}
            ]
        }
    }
}
\begin{document}

\title{OpenRASE: Service Function Chain Emulation}

\author{
\IEEEauthorblockN{Theviyanthan Krishnamohan}
\IEEEauthorblockA{\textit{School of Computing Science} \\
\textit{University of Glasgow}\\
Glasgow, United Kingdom \\
theviyanthan.krishnamohan@glasgow.ac.uk}
\and
\IEEEauthorblockN{Paul Harvey}
\IEEEauthorblockA{\textit{School of Computing Science} \\
\textit{University of Glasgow}\\
Glasgow, United Kingdom \\
paul.harvey@glasgow.ac.uk}
}

\maketitle

\begin{abstract}
Service Function Chains (SFCs) are one of the key enablers in providing programmable computer networks, paving the way for network autonomy. However, this also introduces new challenges, such as resource allocation and optimisation related to their operation, requiring new algorithms to address these challenges. Various tools have been used in the literature to evaluate these algorithms. However, these tools suffer from inaccuracy, low fidelity, unscalability, inflexibility, or additional code requirements. This paper introduces an emulator based on  Mininet and Docker for SFCs called \textit{OpenRASE}. The goal of OpenRASE is to enable the exploration of resource allocation algorithms for SFCs in a dynamic setting, allowing real CPU usage and latency to be measured. We describe the design and implementation of OpenRASE and discuss its characteristics. We also experimentally evaluate two different algorithms to address the SFC resource allocation challenge, including an online Genetic Algorithm, using OpenRASE to show its effectiveness and practicality for dynamic network conditions. 
\end{abstract}


\section{Introduction}
Computer networks are an essential part of everyday life, playing a major role in our economy, education, safety, and especially during COVID-19, our mental health \cite{Imai2020TowardsNetwork}. As a result, it is essential to ensure their continued high-quality operation. Enabling our networks to operate more autonomously is one path to support this, which can lead to cost savings, quality of service increases, more energy-efficient operation, and scaling beyond current human capacity \cite{10255468}. 

One of the key technological enablers on this path is network virtualisation via Service Function Chains (SFCs). SFCs deploy network functions, such as firewalls, virtually on servers,  and establish network links between them using Software Defined Networking (SDN). In effect, SFCs introduce a virtual service overlay over a physical network. 

Despite their flexibility, SFCs also introduce new challenges \cite{GilHerrera2016}, such as how to compose the structure of an SFC (Chain Composition problem), how to allocate network, computational, and storage resources to deploy an SFC's network functions (Forwarding Graph Embedding problem), and how to schedule the SFC's network functions (Scheduling problem). These challenges are collectively termed the Network Function Virtualisation Resource Allocation (NFV-RA) problem \cite{GilHerrera2016}. 

From the literature, there are several approaches to generating solutions to the NFV-RA problem (referred to as \textit{solver} in the remainder of the paper), as well as several different tools and techniques to evaluate these solutions (Section \ref{existing_tools}). Unfortunately, currently available evaluation tools either lack sufficient fidelity, especially for dynamic network environments, or do not support NFV-RA evaluation natively. This limits their effectiveness in providing a robust evaluation of NFV-RA solvers. This is particularly challenging when trying to address the NFV-RA problem for real network environments that exhibit runtime dynamic behaviour (Section \ref{online}).

To address this challenge, we introduce \textbf{Open} \textbf{R}esource \textbf{A}llocation to \textbf{S}FC \textbf{E}mulator (OpenRASE). OpenRASE (Section \ref{features}) is an emulation environment for SFCs based on the Mininet network emulator and Docker virtualisation. To the best of the authors' knowledge, OpenRASE is the first emulator specifically designed to support emulated NFV-RA experiments. To characterise the capabilities of OpenRASE, we compare it against ALEVIN, the most popular tool in the literature for developing solutions to the NFV-RA problem, by experimentally using an ALEVIN solver in OpenRASE. We also demonstrate experimental evaluation of a Genetic Algorithm-based solver to show the versatility of OpenRASE.

\section{Related Work}\label{existing_tools}
A review of the literature related to the NFV-RA problem domain between 2014 and 2025 was performed to examine the tools used to evaluate the NFV-RA algorithms. The tools found in the literature can be categorised as simulators, emulators, testbeds, and production-grade virtualisation tools.

\textit{Simulators} are software programs that model reality at a reduced level of accuracy. They provide a faster, scalable,  flexible and cost-effective way of testing solutions compared to emulators and testbeds. However, they lack fidelity and, as a result, may not provide an accurate evaluation \cite{Fahmy2023SimulatorsWSNs}. 

\textit{Testbeds} provide a controlled physical environment based on the operational or deployment environment and can provide a very high-fidelity evaluation. However, they are expensive, require maintenance, and are less flexible and scalable compared to simulation \cite{Fahmy2023SimulatorsWSNs}.

\textit{Emulators} exist in the continuum between simulators and testbeds \cite{Fahmy2023SimulatorsWSNs}. Instead of modelling or abstracting reality, they duplicate reality, providing higher fidelity than simulators and higher flexibility and scalability than testbeds. Like testbeds, emulators can express uncertainty in results, leading to more realistic and variable behaviour during evaluation cycles. This often comes at the cost of higher execution times compared to simulators. 

\textit{Production-grade virtualisation tools (PGVTs)}, such as OpenStack and Kubernetes, have also been used to evaluate solutions to the NFV-RA problem. Given these are virtualisation tools, it can be argued that they are similar to emulators in that they are more flexible and scalable than testbeds, while offering higher fidelity than simulators. However, they do not support NFV-RA scenarios natively, thus requiring additional code. 

A summary of the tools in the literature can be found in Table \ref{tab1}. It can be seen that most works use simulators in developing NFV-RA algorithms. The use of a simulator like ALEVIN, the most popular tool in the literature, abstracts away challenges associated with dynamic runtime environments, also known as \textit{online experimentation} (Section \ref{online}), such as time consumption and an uncertain environment. As a result, solutions to the NFV-RA problem may not operate effectively after deployment in production systems.
\begin{table}[htbp]
\caption{Summary of the simulators, emulators, testbeds, and Production-Grade Virtualisation Tools found in the literature.}
\begin{center}
\begin{tabular}{>{\centering\arraybackslash}p{0.2\linewidth}>{\centering\arraybackslash}p{0.2\linewidth}>{\centering\arraybackslash}p{0.2\linewidth}>{\centering\arraybackslash}p{0.18\linewidth}}
\hline
\textbf{Name}& \textbf{Type}& \textbf{Supports NFV-RA}& \textbf{Frequency}\\
\hline
ALEVIN& Simulator& Forwarding Graph Embedding Only& 12\\
Mininet& Emulator& No& 7 \\
CloudSim& Simulator& No& 2 \\
OpenStack& PGVT& No& 5 \\
Kubernetes& PGVT& No& 2 \\
OpenNetVM& PGVT& No& 2 \\
OMNet++& Simulator& No& 2 \\
CREATE-NET& Testbed& No& 1 \\
5GinFIRE& Testbed& No& 1 \\
RPi platform& Testbed& No& 1 \\
\hline
\end{tabular}
\label{tab1}
\end{center}
\end{table}

At the same time, as experiments often require frequent changes to the underlying substrate network, testbeds present challenges of scalability and flexibility. Emulators and PGVTs are flexible and scalable and do not abstract away the challenges associated with dynamic runtime environments. However, they require additional code to support evaluating NFV-RA algorithms. 

\section{OpenRASE}\label{features}
To enable accurate online NFV-RA experiments, we now introduce OpenRASE. OpenRASE is an emulator developed to evaluate NFV-RA solvers and is available online\footnote{\url{https://github.com/Project-Kelvin/OpenRASE}}.

\subsection{Design Principles}
\subsubsection{Standards Compliant}
OpenRASE is based on the European Telecommunication Standard Institute (ETSI)'s \cite{NFV2014} NFV architecture and Internet Engineering Task Force (IETF)'s \cite{Halpern2015} SFC architecture. The NFV architecture provides a framework for deploying and managing Virtual Network Functions (VNFs), while the SFC architecture is used to create and maintain SFCs on networks. 

\subsubsection{Python, Mininet, and Docker-based}
OpenRASE is built on top of Containernet \cite{Peuster2017}, a fork of Mininet that enables Docker containers to be used as Mininet hosts. It uses these Docker containers to emulate servers (also referred to as hosts in this paper) and Mininet to establish network connections among Docker containers. Mininet was chosen because it is a popular emulator for performing networking experiments and uses Open vSwitches, providing accurate evaluation of NFV-RA algorithms. 

OpenRASE is written in Python as Mininet provides Python-based APIs\footnote{\url{https://mininet.org/api/hierarchy.html}} and it is a popular high-level language with a simple syntax. Ryu\footnote{\url{https://github.com/faucetsdn/ryu}} is used as the SDN controller since Mininet supports it out of the box. 

\subsubsection{Flexibility}\label{flex}
OpenRASE supports any network function that can be Dockerised. Docker containers that emulate servers run Docker-in-Docker (DinD) images. DinD images allow Docker containers to be nested within Docker containers. Consequently, VNFs can be deployed as Docker containers within DinD containers that emulate servers. This allows OpenRASE to emulate servers running multiple VNFs. 

\subsubsection{Programmable Resource Configuration}
OpenRASE allows configuring the number of CPUs and the amount of memory available to hosts, and the bandwidth of the network links connecting switches and hosts. This enables OpenRASE to realistically mimic real-life networks.

\subsubsection{HTTP-based Traffic}
OpenRASE uses HTTP traffic to run experiments. HTTP was chosen because it can be manipulated easily, as it is an application-layer protocol. This allows traffic to be generated and measured using high-level libraries. The use of HTTP also allows SFC encapsulation to be implemented using an HTTP header attribute. This involves adding the SFC identifier and the sequence of VNFs to traverse to the header. 

\subsubsection{Telemetry collection}
OpenRASE supports the collection of telemetry data from both hosts and network switches. Docker APIs are used to collect the CPU and memory usage data of hosts. The sFlow-RT\footnote{\url{https://sflow-rt.com/}} analytics engine is used to collect bandwidth usage data from the switches. This data can be utilised by NFV-RA algorithms to provide more optimal solutions.

\subsection{SFC Emulation Environment in OpenRASE}\label{architecture}
\begin{figure}[t]
\centerline{\includegraphics[width=0.5\textwidth]{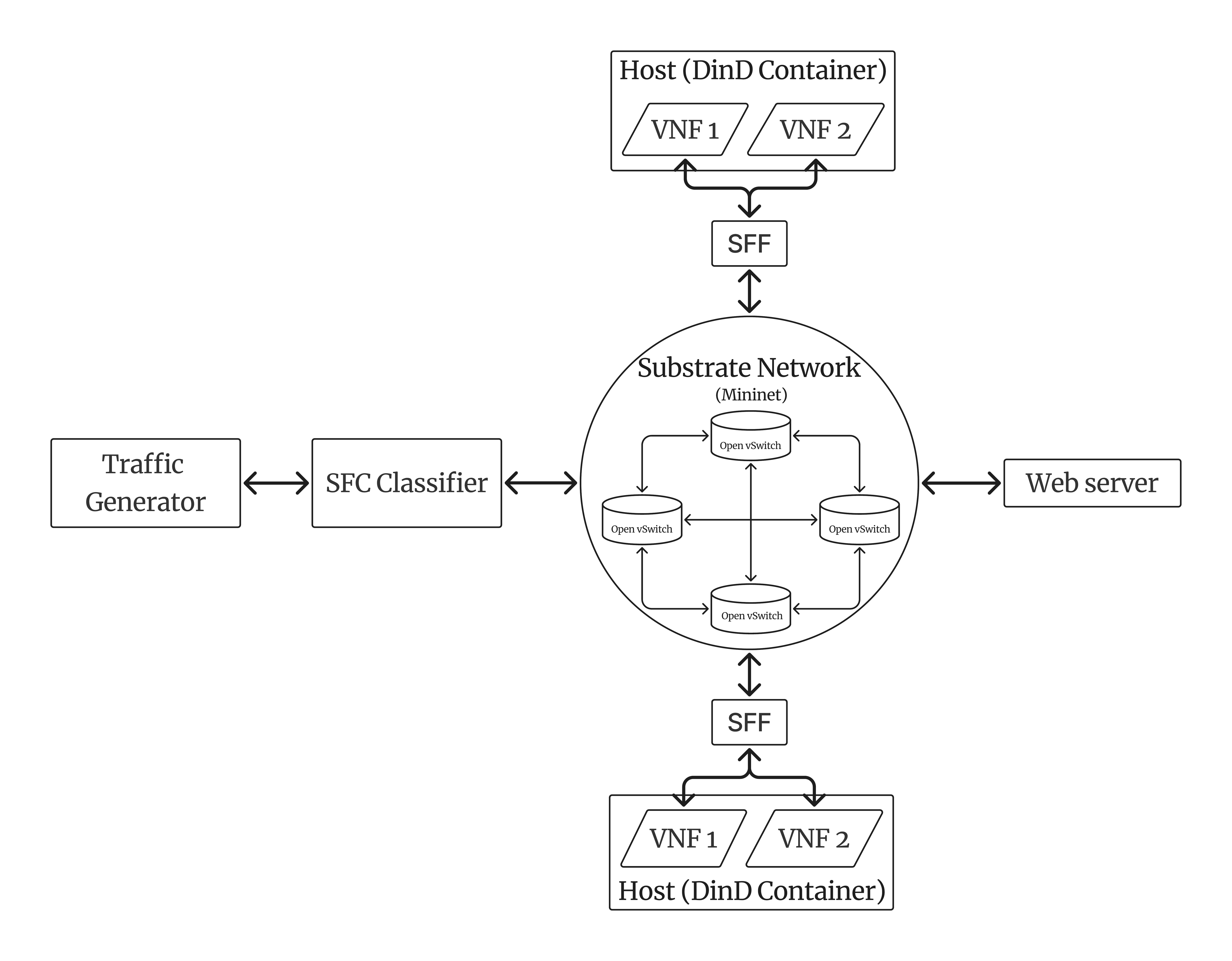}}
\caption{SFC Emulation Environment Provided by OpenRASE}
\label{architecture_fig}
\end{figure}

The SFC emulation environment provided by OpenRASE is based on the IETF SFC architecture  \cite{Halpern2015} and is composed of 6 main elements: the Traffic Generator, the SFC Classifier (SFCC), the Service Function Forwarder (SFF), the hosts, the substrate network, and the web server. Fig.\ \ref{architecture_fig} shows these elements and the interaction between them. 

\subsubsection{Hosts}
The hosts emulate the servers where VNFs are deployed. In OpenRASE, hosts are DinD containers that run VNFs as containers within them. Each host is connected to a Service Function Forwarder (SFF) (see Section \ref{sff}).

\subsubsection{Web Server}\label{server}
The web server is the endpoint that the traffic generated by the Traffic Generator eventually reaches after traversing an SFC. The server hosts a REST GET API endpoint. The web server is hosted in a dedicated host.

\subsubsection{Traffic Generator}\label{tg}
The Traffic Generator generates HTTP traffic by sending REST API requests tagged with an SFC ID. The SFC ID uniquely identifies each SFC deployed on the network. The Traffic Generator generates GET requests using Grafana K6\footnote{\url{https://github.com/grafana/k6}}, an open-source tool used to load test APIs, to the web server. This tool enables the number of requests per user-defined unit of time to be adjusted to mimic real traffic patterns.

\subsubsection{SFC Classifier (SFCC)} \label{sfcc}
The SFCC classifies ingress traffic and routes it to appropriate SFCs based on the SFC ID found in the HTTP header. It is implemented as an application executing on a Docker host. 
Once an SFC is identified, the SFCC forwards it to the SFF  (Section \ref{sff}) that corresponds to the host that hosts the first VNF in the SFC. 

\subsubsection{Service Function Forwarder (SFF)}\label{sff}
An SFF is responsible for forwarding the traffic to the appropriate VNFs running within the host. After the traffic is processed by the VNF, the SFF is also responsible for forwarding the traffic to the SFF corresponding to the host that contains the next VNF in the SFC. The SFF is an app that runs on a dedicated Docker container that is connected to its host.

\subsubsection{Substrate Network}
The substrate network is a Mininet environment that emulates a physical network. As Mininet uses Open vSwitches, the network can be configured using the OpenFlow SDN protocol. The Ryu SDN controller is used to manipulate the flow tables of the switches so that traffic to an SFC can be directed through the switches.

\subsection{OpenRASE Software Design}\label{design}
\begin{figure}[t]
    \centering
    \includegraphics[width=0.5\textwidth]{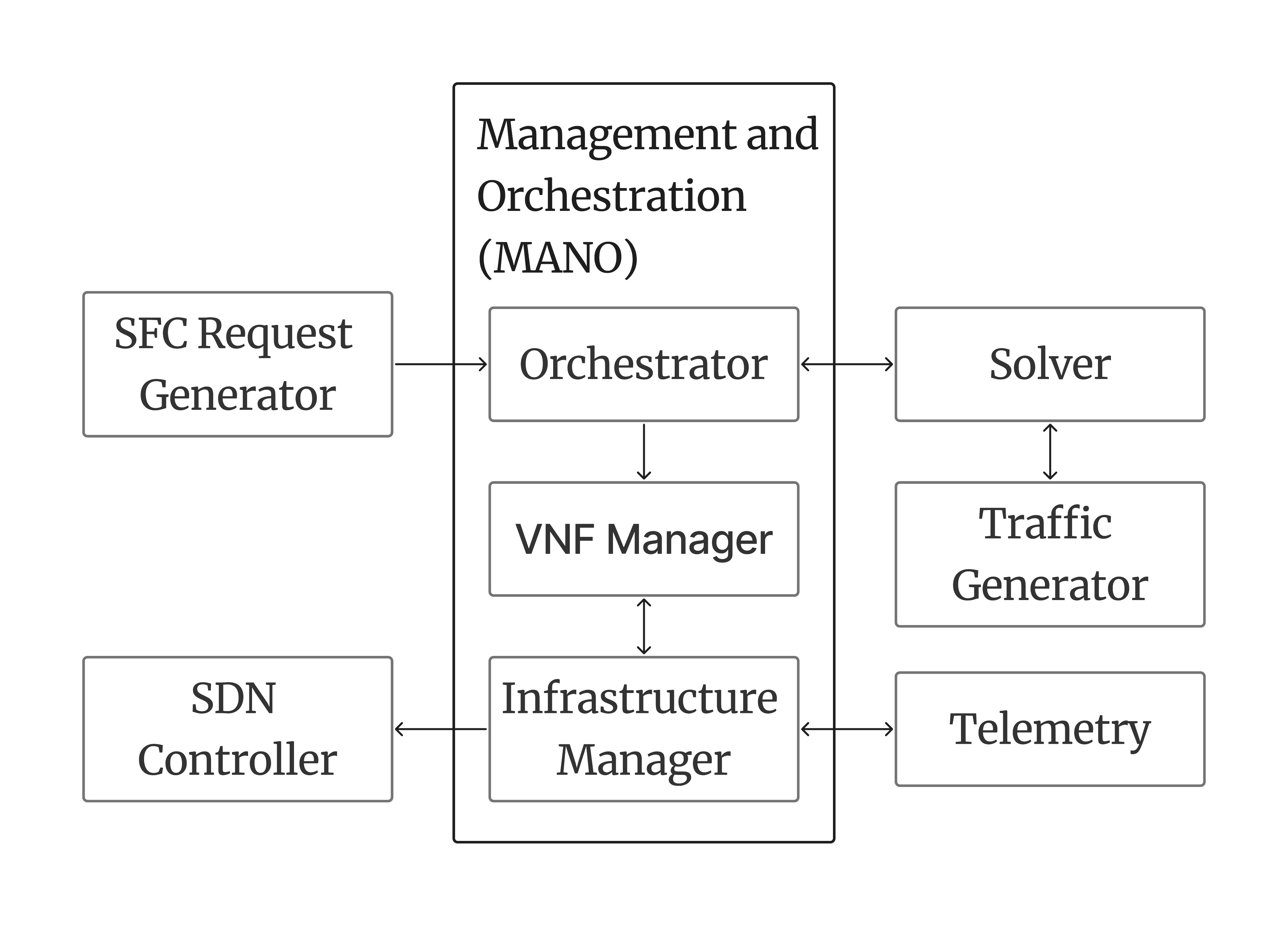}
    \caption{OpenRASE Architecture}
    \label{fig:design}
\end{figure}

OpenRASE uses a modular software design based on the ETSI NFV architectural framework I \cite{NFV2014}. It consists of 8 modules: the Orchestrator, the VNF Manager, the Infrastructure Manager, the SDN Controller, the Telemetry, the Traffic Generator, the SFC Request Generator, and the Solver, as shown in Fig.\ \ref{fig:design}. The Orchestrator, the VNF Manager, and the Infrastructure Manager together form the Management and Orchestration (MANO) module.

\subsubsection{SFC Request Generator}
This module is responsible for generating SFC Requests (SFCRs). These requests are sent to the Orchestrator, which in turn calls the Solver module to generate optimal deployment schemes.

\subsubsection{Solver}\label{solver}
The Solver module is an abstract module that can be used to implement an approach to solving the NFV-RA problem (solver), such as the Simple Dijkstra algorithm (Section \ref{evaluation}) or Genetic Algorithm (Section \ref{online}). 

\subsubsection{Orchestrator}
This module coordinates the communication between modules within and outwith the MANO module. It is also responsible for configuring the emulation environment to a specified deployment scheme by orchestrating VNFs through the VNF Manager and deploying the network topology through the Infrastructure Manager. This module also provides the Solver module with the necessary telemetry data (Section \ref{subsub:telemetry}).

\subsubsection{VNF Manager}
The VNF Manager module deploys and removes VNFs to and from hosts.

\subsubsection{Infrastructure Manager}
This module is responsible for configuring the substrate network and assigning and managing the IP addresses of the hosts. The Infrastructure Manager uses Mininet's Python APIs to create hosts and switches. It also coordinates with the SDN Controller to configure the switches with the necessary flows. The Infrastructure Manager also creates and manages the web server and the SFC Classifier. 

\subsubsection{SDN Controller}
The SDN Controller uses the Ryu SDN controller's REST APIs to install and manage flows on the Open vSwitches, as well as assign and manage the IP addresses of the switches.

\subsubsection{Telemetry}\label{subsub:telemetry}
This module is responsible for collecting telemetry data from hosts and switches. Docker's APIs are used to collect telemetry data of hosts, while sFlow-RT is used to collect telemetry data of the switches. 

\subsubsection{Traffic Generator}
This module orchestrates and manages the Traffic Generator component. As discussed in Section \ref{tg}, the Traffic Generator component is based on Grafana K6. This module initiates traffic to SFCs as soon as they are deployed and collects traffic latency data.

\section{ALEVIN vs. OpenRASE}\label{feature_comp}
To place OpenRASE in context, we compare it against ALEVIN, the most popular tool in the literature for developing solutions to the NFV-RA problem (Section \ref{existing_tools}). Table \ref{comp_sum} shows a comparison of the features of OpenRASE and ALEVIN.

By emulating an SFC environment and operation more accurately and flexibly, it can be seen that OpenRASE offers equivalent and additional evaluation capabilities to ALEVIN to explore the NFV-RA problem. This higher fidelity comes at the cost of higher resource requirements compared to ALEVIN, especially for more complex topologies. 

A key difference between ALEVIN and OpenRASE is that OpenRASE uses HTTP traffic requests to drive the emulation of the VNFs. ALEVIN does not contain the concept of traffic.

The resource requirements of VNFs, such as their CPU and memory demands, which are important to developing NFV-RA algorithms, are user-specified in ALEVIN. However, in OpenRASE, these requirements can be obtained by measuring resource usage during experiments, bringing the evaluation of these algorithms closer to reality. In ALEVIN, the resource consumption of VNFs remains static, whereas in OpenRASE, resource consumption changes in relation to the amount of traffic. This facilitates the development of algorithms that can respond to \textit{dynamic} traffic patterns.

Conversely, ALEVIN is a lightweight tool compared to OpenRASE. OpenRASE has a high CPU and memory demand as it uses Docker containers and Open vSwitches. Furthermore, new VNFs with different resource demands can be easily added to ALEVIN, as they are simulated. This only requires static values related to the VNFs. Doing the same in OpenRASE requires additional effort as real VNFs must be implemented and integrated with the emulator.  

\begin{table}
\centering
\caption{A comparison between ALEVIN and OpenRASE}
\label{comp_sum}
\begin{tabular}{>{\centering\arraybackslash}p{0.20\linewidth}>{\centering\arraybackslash}p{0.30\linewidth}>{\centering\arraybackslash}p{0.33\linewidth}} \hline  
 \textbf{Factor}& \textbf{ALEVIN} & \textbf{OpenRASE} \\ \hline
Experiments & Experiments are simulated and do not involve traffic generation.& Experiments are emulated using HTTP traffic.\\ 
Hosts & Simulated. A host is represented via a software model.& Docker containers as hosts.\\ 
Switches & Simulated. A switch is represented via a software model.& Open vSwitches using Mininet.\\ 
VNFs & Simulated and abstract units.& 7 distinct VNFs. Container-based deployable code. \\ 
Per-VNF resource requirements & Arbitrary, static, and user-specified.& Automatically extracted from VNFs across varying traffic demand by benchmarking.\\ 
SFC Deployment& Simulated & Emulated deployment using Docker containers and Mininet.\\ 
Language & Java; there is no VNF logic implementation & Python;  VNFs can use any technology/language.\\ 
Resource Requirements & Low: can run on a laptop. & High:  Experimenting with complex topologies requires servers with adequate resources(Section \ref{setup}).\\ 
Incorporating new VNF resource types (e.g., memory, bandwidth, disk space)& Can be added since they are simulated. & Requires code-level changes. Support for new resource types may be limited by Docker and Mininet.\\ \hline

\end{tabular}

\end{table}

\section{Evaluation}\label{evaluation}
To demonstrate the effectiveness of OpenRASE in the evaluation of NFV-RA solvers, we evaluate both OpenRASE and ALEVIN using the Simple Dijkstra algorithm present in ALEVIN\footnote{\url{https://sourceforge.net/p/alevin/svn/HEAD/tree/trunk/src/tests/algorithms/SimpleDijkstraAlgorithmTest.java}}. Each experiment receives a specified number of SFCRs and tries to allocate resources to the SFCs such that the acceptance ratio of SFCRs (the number of SFCRs that can be deployed as a fraction of the total SFCRs) is maximised.

The objective of this evaluation was to show that both ALEVIN and OpenRASE would produce the same acceptance ratio when all conditions are equal, while demonstrating the ability of OpenRASE to generate real traffic and measure the latency and CPU usage of hosts. Latency is the round-trip time that an HTTP request takes to traverse an SFC from the Traffic Generator to the web server and back, and is a significant consideration for real systems. As ALEVIN does not support traffic generation, latency measurement is not possible.

\subsection{Setup}\label{setup}
The evaluation was carried out on a virtual machine running Ubuntu 20.04.6 on a QEMU hypervisor with 64 cores of Intel Xeon Gold 6240R CPUs having a clock speed of 2.4 GHz, and 64GB of RAM. 

\subsection{Experiments}\label{experiments}
Eight experiments were created to compare OpenRASE against ALEVIN \footnote{Experiment data is available here: \url{https://github.com/Project-Kelvin/open-research-data/tree/main/openrase_softcom/experiment_server_03_06_2024}}. The acceptance ratio was recorded for each experiment across both tools. In addition, the deployment time, the CPU usage of hosts, and the average latency of SFCs were measured in OpenRASE. These measurements are not possible in ALEVIN. In each experiment, the number of CPUs per host and the submitted SFCRs were varied.

The number of CPUs available in each host was set to 2 or 4. The number of SFCRs was adjusted by creating multiple instances of each of the 4 unique SFCRs, as shown in Table \ref{tab:results}.

\subsection{NFV-RA Solver}
The NFV-RA solver Simple Dijkstra algorithm, available by default in ALEVIN, was implemented in OpenRASE using the Solver module. Simple Dijkstra deploys VNFs on servers in a greedy manner and then uses the Dijkstra algorithm to link the VNFs. 

\section{Results}\label{results}
\begin{table*}[h]
    \centering
\caption{Evaluation results}
\label{tab:results}
    \begin{tabular}{>{\centering\arraybackslash}p{0.10\linewidth}>
    {\centering\arraybackslash}p{0.10\linewidth}>
    {\centering\arraybackslash}p{0.10\linewidth}>
    {\centering\arraybackslash}p{0.10\linewidth}>
    {\centering\arraybackslash}p{0.10\linewidth}>{\centering\arraybackslash}p{0.10\linewidth}>{\centering\arraybackslash}p{0.20\linewidth}} \hline
 \multirow{3}{\linewidth}{\textbf{Experiment}} & \multirow{3}{\linewidth}{\textbf{CPUs per host}} &  \multirow{3}{\linewidth}{\textbf{SFCRs Duplicates}}&  \multirow{3}{\linewidth}{\textbf{Submitted SFCRs}} & \multicolumn{2}{c}{\textbf{Acceptance Ratio}}&\textbf{Deployment Time (s)}\\[5pt] \cline{5-7}        
     & & &&\textbf{OpenRASE}&\textbf{ALEVIN}&\textbf{OpenRASE}\\ \hline 
         1 & 2& 1&4& 1& 1& 34.02\\ 
         2 & 2& 2&8& 1& 1& 44.43\\ 
         3 & 2& 4&16& 1& 1& 75.76\\ 
         4 & 2& 8&32& 0.75& 0.75& 90.81\\  
         5 & 4& 1&4& 1& 1& 28.58\\  
         6 & 4& 2&8& 1& 1& 34.39\\  
         7 & 4& 4&16& 1& 1& 55.04\\  
         8 & 4& 8&32& 1& 1& 110.69\\ \hline
    \end{tabular} 
\end{table*}
Both ALEVIN and OpenRASE produced the same acceptance ratios for all experiments, as shown in Table \ref{tab:results}. This demonstrates OpenRASE's ability to produce the same outcome as ALEVIN.

Experiments on ALEVIN on average took less than 10ms each to complete, whereas an experiment on OpenRASE took approximately 10 minutes on average. This was expected as it is a tradeoff between simulation and emulation.

The deployment time and latency of each SFC and the CPU usage of hosts were recorded in OpenRASE. The deployment time is the time taken to deploy the VNFs on hosts and establish links between them. This metric, along with the latency and CPU usage, is irrelevant for ALEVIN. The ability to deploy SFCs, generate real traffic, and measure the latencies of SFCs along with the CPU usage of hosts is a significant advantage that OpenRASE offers over ALEVIN.

\begin{figure}
    \centering
    \includegraphics[width=1\linewidth]{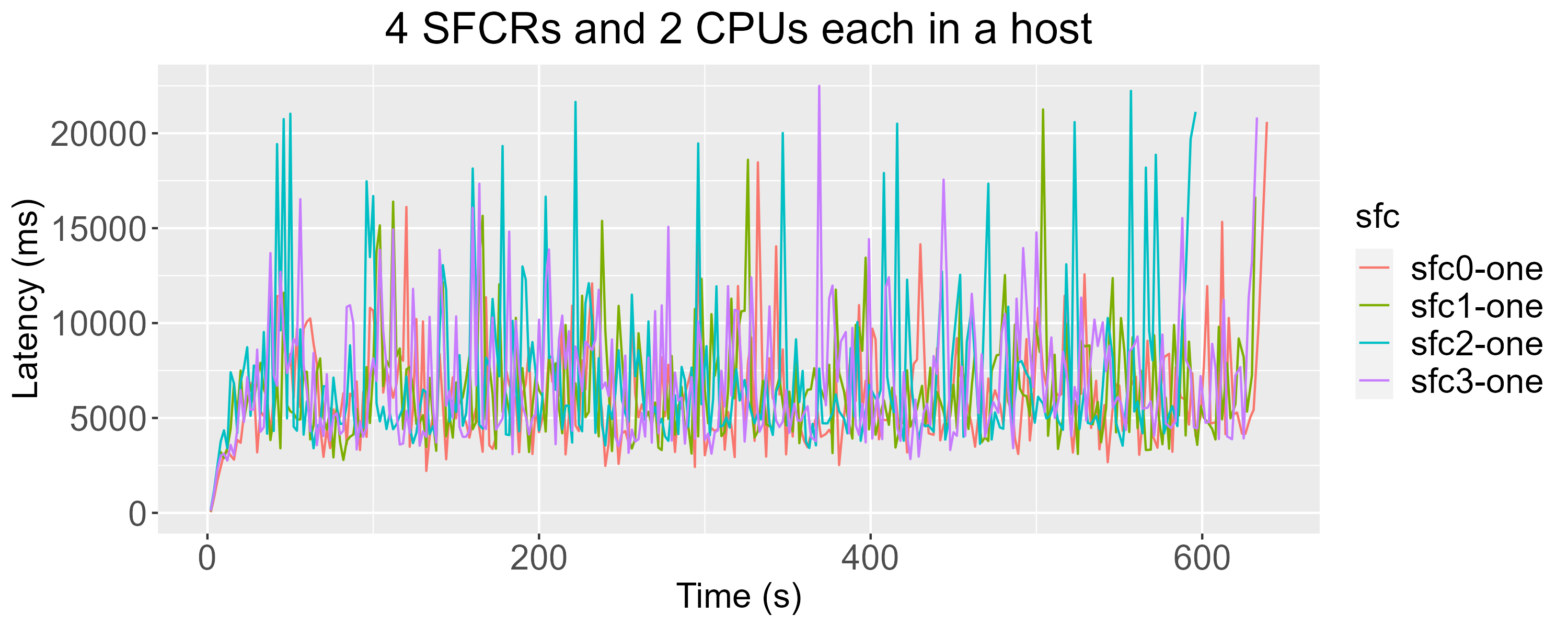}
    \caption{Binned response latency of SFCs over time in Experiment 1}
    \label{fig:latency}
\end{figure}

\begin{figure}
    \centering
    \includegraphics[width=1\linewidth]{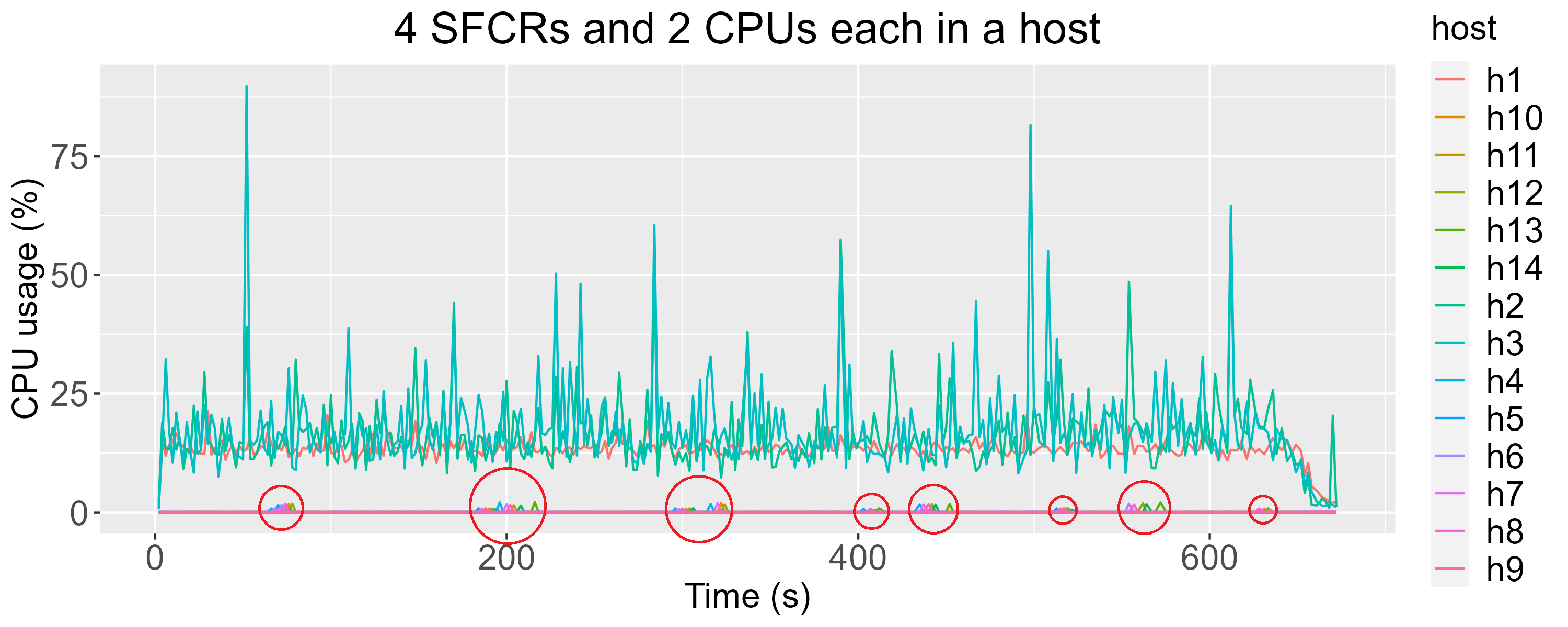}
    \caption{CPU usage in each host over time in Experiment 1. Sporadic spikes in CPU usage in idle hosts are highlighted by red circles. }
    \label{fig:cpu_usage}
\end{figure}

Table \ref{tab:results} shows the acceptance ratio and deployment time of OpenRASE for each experiment. The binned latency of each SFC and the CPU usage of each host over time in Experiment 1 are given in Fig.\ \ref{fig:latency} and Fig.\ \ref{fig:cpu_usage}, respectively. 


\subsection{Discussion}\label{discussion}
The acceptance ratios obtained from experiments run on ALEVIN and OpenRASE were identical, as shown in Table \ref{tab:results}. Thus, when the same algorithm is used, OpenRASE and ALEVIN produce the same results. However, this is a static metric, unlike the realistic, dynamic operational metrics that OpenRASE can measure. 

CPU usage and traffic latency are dynamic operational metrics. Variations in CPU usage and latency across time can be observed in experiments run on OpenRASE. Small spikes in CPU usage can be observed in idle hosts as shown by red circles in Fig.\ \ref{fig:cpu_usage}. Such uncertainties are an advantage that emulators provide over simulators when it comes to more accurately replicating the dynamic environments of networks, as they enable the adaptability of algorithms to uncertainties to be evaluated. 

An experiment carried out on OpenRASE took nearly 10 minutes to complete. This is because real HTTP traffic was used for evaluation. This high time consumption becomes challenging when the number of experiments increases. However, this also provides an environment closer to reality, as an experiment on a real network is going to consume just as much time. 

An approach that solves the NFV-RA problem by evolving solutions via online experimentation (Section \ref{online}) has to address the challenges of high time consumption and an uncertain environment to be practical enough to be applied in real networks. OpenRASE, by posing these challenges, enables such approaches to be evaluated accurately. 

\section{An Experimental Online Evolution Using OpenRASE}\label{online}
As described previously, OpenRASE enables the emulation of an SFC environment that dynamically changes over time. Unlike the more static approaches, OpenRASE enables solvers to be created and validated in a dynamic network context. This is known as \textit{online experimentation}~\cite{10255468} and has been identified as a key principle in achieving fully autonomous network operation.  

\begin{figure}
    \centering
    \includegraphics[width=1.0\linewidth]{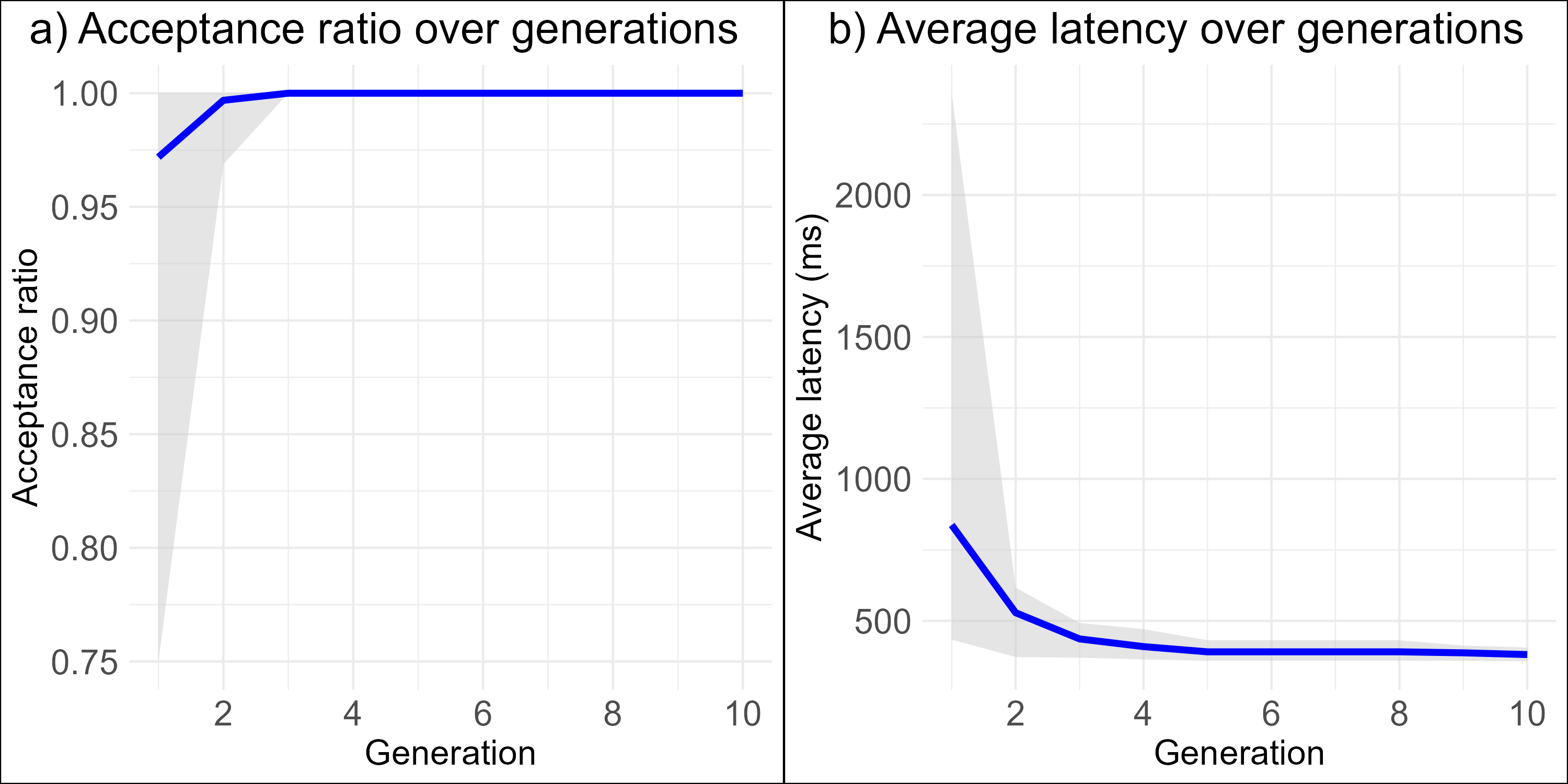}
     \caption{Two plots showing the average acceptance ratio (a) and the average latency in ms (b) of the population per generation. The blue line shows the average; the shaded area shows the range.}
    \label{fig:ar_lat_plot}
    \label{fig:enter-label}
\end{figure}
One approach to realise network autonomy is evolutionary networks \cite{10255468}. An evolutionary approach adapts and improves solutions gradually at runtime to solve problems, such as NFV-RA, optimally. An experimental online evolutionary approach combines both online experimentation and evolution. Such an approach implements a solution in a network, evaluates it, and iteratively evolves it to optimally solve problems.

As a preliminary study to demonstrate the usefulness of OpenRASE in performing online experimentation and as a path to network autonomy, an \textit{online} algorithm was developed to optimise both the acceptance ratio \textit{and} the average latency of an SFC deployment. This algorithm uses a Genetic Algorithm to decide the hosts in which VNFs would be deployed and the Dijkstra algorithm to route traffic between these VNFs. Accordingly, potential solutions are generated (population), evaluated on OpenRASE, and iteratively evolved using the Genetic Algorithm to produce a solution.

The experiment\footnote{Experiment data is available here: \url{https://github.com/Project-Kelvin/open-research-data/tree/main/openrase_softcom/experiment_server_20_07_2024}} converged on an average acceptance ratio of 1 and an average latency of 381.38ms over 10 generations, as shown in Figure \ref{fig:ar_lat_plot}. The average latency was a value obtained by measuring real traffic and was not mathematically computed, as in ALEVIN. This is a significant advantage OpenRASE offers over other similar tools. However, the trade-off is that OpenRASE took nearly 19 hours to complete this experiment. This is impractical as the longer waiting time to find a solution can incur monetary losses. 

Thus, a key area of future work is to reduce the time consumption for emulation and, more generally, continue to explore the use of Genetic Algorithm-based solutions via online experimentation for the NFV-RA problem.

\section{Conclusion}\label{conclusion}
This paper describes OpenRASE, an emulator for Service Function Chains (SFC) to enable the evaluation of solvers for the Network Function Virtualisation Resource Allocation (NFV-RA) problem in dynamic environments. To the best of our knowledge, this is the first emulator for this problem. We discuss the different tools used in the literature to evaluate NFV-RA solvers and note a gap in relation to emulators. We compare OpenRASE with ALEVIN, the most popular tool in the literature, presenting an experimental comparison across both static and dynamic solvers and network environments. Compared to ALEVIN, and at the expense of higher execution and deployment times, OpenRASE produces equivalent algorithmic results for static contexts. We also experimentally show that OpenRASE supports the development of online Genetic Algorithms for the NFV-RA problem in dynamic contexts while considering latency, which ALEVIN can not. 

\printbibliography
\end{document}